\documentclass[preprint,showpacs,preprintnumbers,amsmath,amssymb,natbib]{revtex4}

\usepackage{graphicx}
\usepackage{epsfig}		
\usepackage{dcolumn}
\usepackage{bm}

\renewcommand{\baselinestretch}{1.4}
\def\0{\mbox{\tiny $0$}}
\def\1{\mbox{\tiny $1$}}
\def\2{\mbox{\tiny $2$}}
\def\3{\mbox{\tiny $3$}}
\def\4{\mbox{\tiny $4$}}
\def\5{\mbox{\tiny $5$}}
\def\6{\mbox{\tiny $6$}}
\def\7{\mbox{\tiny $7$}}
\def\8{\mbox{\tiny $8$}}
\def\9{\mbox{\tiny $9$}}

\def\f14{\mbox{\tiny $\frac{1}{4}$}}

\def\bb#1{\mbox{\footnotesize $(#1)$}}

\begin{document}
%
\title{Quantum measurement schemes related to flavor-weighted energies}
\author{A. E. Bernardini}
\email{alexeb@ufscar.br}
\affiliation{Departamento de F\'{\i}sica, Universidade Federal de S\~ao Carlos, PO Box 676, 13565-905, S\~ao Carlos, SP, Brasil}

\date{\today}

\begin{abstract}
The framework of the generalized theory of quantum measurement provides some theoretical tools for computing the von-Neumann entropy correlated to flavor associated energies.
It allows for relating flavor associated energies to {\em non-selective} ({\em selective}) quantum measurement schemes.
Reporting about the density matrix theory for a composite quantum system of flavor eigenstates, we introduce the idea of flavor-{\em weighted} energies.
It provides us with the right correlation between the energies of flavor eigenstates and their measurement probabilities.
In addition, the apparent ambiguities which follow from computing flavor-{\em averaged} energies are suppressed.
As a final issue, the connection of such flavor associated energies with the expressions for neutrino effective mass values is investigated.
It is straightforwardly verified that cosmological background neutrino energy densities could be obtained from the coherent superposition of mass eigenstates.
Our results show that the {\em non-selective} measurement scheme for obtaining flavor-{\em weighted} energies is consistent with the predictions from the single-particle quantum mechanics.
\end{abstract}

\pacs{14.60.Pq, 03.65.Ta, 03.65.Yz}

\keywords{quantum measurement, flavor oscillation, density matrix, cosmology}

\maketitle

\newpage
\section{Introduction}

Although quantum mechanics is an intrinsically probabilistic theory, its inherent application of probabilistic concepts is quite different from that of a classical theory.
In particular, in the framework of the quantum mechanics for composite systems, the density matrix is the analogue to the probability distribution of position and momentum in classical statistical mechanics.
Such a statistical description through density matrices is required when one considers either an ensemble of systems, or a composite quantum system defined when its preparation history is uncertain and one does not know if it is a pure quantum state or a statistical mixture.
That should be the case of an ensemble of neutrino flavor eigenstates, for instance, in the cosmological scenario.

The description of measurements performed on a composite quantum system provides an important tool to debug the procedures for computing the averaged energy densities of cosmological neutrino flavor eigenstates.
We shall demonstrate that it should result in the appropriate relation between the cosmological neutrino background energy densities and neutrino mass values.
In fact, since one asserts that the generalized theory of quantum measurement is based on the notions of operations and effects, the correlation between measurable flavor neutrino energies and flavor eigenstates, at least from the formal point of view, deserves some specific attention \cite{Giu91,Zra98,Zub98,Giu98}.

Departing from the standard formulation of composite quantum systems, we present the fundamentals of the physical significance of the density matrix theory in computing flavor probabilities and flavor-{\em averaged} energies.
To overcome the ambiguities and misunderstandings that arise when flavor-{\em averaged} energies are defined, we discuss the idea of flavor-{\em weighted} energies mathematically correlated to flavor probabilities.
Reporting about the generalized theory of quantum measurements, it is demonstrated that one can depict the {\em averaged} and {\em weighted} energy definitions from the idea of {\em selective} and {\em non-selective} quantum measurements.
It is shown that such {\em weighted} energies based on some relations with statistical weights are more convenient in describing certain properties of composite quantum systems strictly related to flavor quantum numbers.

Flavor energy ``measurements'' or ``projections'' are therefore potentially subject to imprecise definitions.
Obviously it reflects the dynamics of quantum systems being driven by a diagonal Hamiltonian in the mass eigenstate basis.
In this case, the interpretation of the von-Neuman entropy for {\em selective} and {\em non-selective} quantum measurements is helpful in verifying correlations with flavor oscillation probabilities.
It is performed by assuming that the von-Neumann-L\"{u}ders projection postulate introduces the concepts of {\em selective} and {\em non-selective} measurements \cite{Breuer}.
It complies with the fundamentals of the generalized theory of quantum measurements \cite{Breuer,10,20,30} developed from the extended idea of a {\em positive operator-valued measure} that associates with each measurement outcome $\alpha$ a positive operator $M^{\alpha}_{(0)}$.

These concepts can play a relevant role in the fine-tunning of the neutrino mass value predictions.
The experimental procedure for determining the mass of the neutrino using the CMB results is by inferring the transfer function in the matter power spectrum at small scales \cite{Dolgov02}.
The contribution due to massive neutrinos to the closure fraction of cold dark matter at present substantially modifies the matter power spectrum, even for neutrinos behaving like hot dark matter at higher redshifts\cite{Boy10,Ma94,Pas06}.
Determining the fraction of the neutrino energy density at late times is therefore a relevant aspect that has to be included in the procedure for deriving neutrino masses.
In this manuscript, the influence of different flavor energy definitions in obtaining the predictions for cosmological neutrino mass values are discriminated.
In particular, we show how the mass predictions are modified by some explicit dependence on the statistical weights of an ensemble of neutrino flavors.

Our manuscript was organized as follows.
In section II we report about the usual mechanism of flavor oscillations through which one deduces the flavor conversion formulas and the expressions for flavor-{\em averaged} energies.
The concept of flavor-{\em weighted} energy is discussed in section III where it is compared with the previously defined flavor-{\em averaged} energy and with the {\em total} averaged energy inherent to composite quantum systems.
The idea of {\em selective} and {\em non-selective} quantum measurements is introduced in manner to embed the definitions of {\em averaged} and {\em weighted} energies.
In section IV, an extension of such concepts is proposed in order to include a connection to flavor associated von-Neumann entropies.
Our results show that there exists a kind of correlation rate between the flavor-{\em weighted} energy and the von-Neumann entropy changes due to a {\em non-selective} measurement scheme.
Finally, potential implications on the properties of the cosmological neutrino background are discussed in section V, where the connection between the {\em weighted} energies and the cosmological neutrino energy density is established.
We draw our conclusions in section VI.

\section{Flavor oscillations}

The aspects of neutrino flavor oscillations that are relevant to our analysis can be comprehended from a simplified treatment involving just two degrees of freedom.
The time evolution of a quantum system of well-defined flavor quantum numbers described by the state vectors $\nu^{e}$ and $\nu^{\mu}$ respectively related to electron and muon neutrinos is given by
\begin{equation}
\left(\begin{array}{l}\nu^e_{(t)} \\ \nu^{\mu}_{(t)} \end{array}\right) =
U \, \left(\begin{array}{ll} e^{-i\, E_{\1} t} & 0 \\ 0  &  e^{-i\, E_{\2} t} \end{array}\right)\,
\left(\begin{array}{l}\nu_1 \\ \nu_2 \end{array}\right) =
U \, \left(\begin{array}{ll} e^{-i\, E_{\1} t} & 0 \\ 0  &  e^{-i\, E_{\2} t} \end{array}\right)
\, U^{\dagger} \left(\begin{array}{l} \nu^e_{(0)} \\ \nu^{\mu}_{(0)}  \end{array}\right),
\label{101A}
\end{equation}
where $\nu_{\1}$ and $\nu_{\2}$ are the mass eigenstates with well-defined energies, $E_{s} = \sqrt{p^{\2} + m^{\2}_{s}}$, with $s = 1,\, 2$, and the matrix $U$ parameterizes the mixing relation as
\begin{equation}
\left(\begin{array}{l}\nu^e_{(0)} \\ \nu^{\mu}_{(0)} \end{array}\right) =
U \, \left(\begin{array}{l}\nu_1 \\ \nu_2 \end{array}\right) =
\left(\begin{array}{rr} \cos\bb{\theta} & \sin\bb{\theta} \\ -\sin\bb{\theta}  & \cos\bb{\theta} \end{array}\right)
\left(\begin{array}{l}\nu_1 \\ \nu_2 \end{array}\right),
\label{eq00A}
\end{equation}
where $\theta$ is the mixing angle.
Since the Hamiltonian of the system in the mass eigenstate basis can be extracted from Eq.~(\ref{eq00A}) as $H = Diag\{E_{\1},\, E_{\2}\}$, the flavor projection operators can be easily defined as
\begin{equation}
M^{e}_{(t)} = |\nu^{e}_{(t)} \rangle \langle \nu^{e}_{(t)}| =
\left[
\begin{array}{ll}
\cos^{\2}\bb{\theta} & \sin\bb{\theta} \,\cos\bb{\theta} \,e^{-i\,\Delta E\, t}\\
\sin\bb{\theta}\, \cos\bb{\theta} \, e^{i\,\Delta E\, t} &  \sin^{\2}\bb{\theta}
\end{array}
\right]
\label{eq01A}
\end{equation}
and
\begin{equation}
M^{\mu}_{(t)} = |\nu^{\mu}_{(t)} \rangle \langle \nu^{\mu}_{(t)}|  =
\left[
\begin{array}{ll}
\sin^{\2}\bb{\theta} & -\sin\bb{\theta} \,\cos\bb{\theta} \,e^{-i\,\Delta E\, t}\\
-\sin\bb{\theta}\, \cos\bb{\theta} \, e^{i\,\Delta E\, t} &  \cos^{\2}\bb{\theta} \end{array}
\right]
\label{eq01B}
\end{equation}
where $\Delta E = E_{\1} - E_{\2}$ and it can be verified that $M^{e}_{(t)} + M^{\mu}_{(t)}  = \mathbf{1}$.

Thus the temporal evolution of a flavor eigenstate can be described by
\begin{equation}
|\nu^{e,\mu}_{(t)} \rangle = (M^{e}_{(0)} + M^{\mu}_{(0)}) |\nu^{e,\mu}_{(t)} \rangle =
 \langle \nu^{e}_{(0)} |\nu^{e,\mu}_{(t)} \rangle \, |\nu^{e}_{(0)} \rangle
+
\langle \nu^{\mu}_{(0)} |\nu^{e,\mu}_{(t)} \rangle \, |\nu^{\mu}_{(0)} \rangle,
\label{eq01C}
\end{equation}
and the supposedly relevant measurable quantities, or {\em observables}, of the closed quantum system can be summarized by the the flavor-{\em averaged} energies,
\begin{eqnarray}
E^{e,\mu}_{(t)} &=&  \langle \nu^{e,\mu}_{(t)} |H| \nu^{e,\mu}_{(t)}\rangle,
\label{eq02}
\end{eqnarray}
that result in time-independent quantities,
\begin{eqnarray}
E^{e}_{(t)} &=& E^{e}_{(0)} = \bar{E} + (1/2) \Delta E\, \cos\bb{2\theta},\nonumber\\
E^{\mu}_{(t)} &=& E^{\mu}_{(0)} = \bar{E} - (1/2) \Delta E\, \cos\bb{2\theta},
\label{eq02A}
\end{eqnarray}
with $\bar{E} = (1/2)(E_{\1} + E_{\2})$, and by the time-oscillating flavor probabilities,
\begin{eqnarray}
\mathcal{P}_{\alpha\rightarrow\beta} \bb{t} &=& Tr\{M^{\beta}_{(0)}\, M^{\alpha}_{(t)}\}, ~~~~\alpha,\,\beta = e,\, \mu,
\label{eq02BB}
\end{eqnarray}
that result in
\begin{eqnarray}
\mathcal{P}_{e\rightarrow e} \bb{t} &=& P_{\mu\rightarrow\mu} \bb{t} = |\langle \nu^{e}_{(\0)}|\nu^{e}_{(t)}\rangle|^{\2} = |\langle \nu^{\mu}_{(\0)}|\nu^{\mu}_{(t)}\rangle|^{\2}  = 1 - \sin^{\2}\bb{2\theta} \,\sin^{\2}\left(\frac{\Delta E}{2}\, t\right),
\label{eq02B}
\end{eqnarray}
and
\begin{eqnarray}
\mathcal{P}_{e\rightarrow \mu}\bb{t} &=& P_{\mu \rightarrow e} \bb{t}= |\langle \nu^{\mu}_{(\0)}|\nu^{e}_{(t)}\rangle|^{\2}  = |\langle \nu^{e}_{(\0)}|\nu^{\mu}_{(t)}\rangle|^{\2} = \sin^{\2}\bb{2\theta} \,\sin^{\2}\left(\frac{\Delta E}{2}\, t\right),
\label{eq02C}
\end{eqnarray}
that are interpreted as the probabilities of $e(\mu)$-flavor states produced at time $t_{\0}$ be measured as $e(\mu)$-flavor states or be converted into $\mu (e)$-flavor states after a time interval $t - t_{\0} \sim t - 0 \sim t$.

Due to the relation between flavor eigenstates described by Eq.~(\ref{eq01C}), the above definition of flavor-{\em averaged} energies is ambiguous in the sense that eventually $e (\mu)$-flavor states can be partially, or even completely, converted into $\mu (e)$-flavor states.
To be more clear, once the projection of the muon vector state at time $t$, $\nu^{\mu}_{(t)}$, onto the initial ($t_{\0} = 0$) electron vector state, $\nu^{e}_{(0)}$, is not zero, i. e. $|\langle \nu^{\mu}_{(\0)}|\nu^{e}_{(t)}\rangle| \neq 0$, the averaged value computed from Eq.~(\ref{eq01C}) represents an ambiguous and inappropriate definition of flavor associated to energies, given that it is not uniquely correlated with the respective flavor eigenstate.
Obviously one has such a crude definition of flavor energy ``measurements'' or ``projections'' because the time evolution of the system is driven by a diagonal Hamiltonian in the mass eigenstate basis.

Such an ambiguity has stimulated some non-standard analysis of the cosmological background neutrinos as a coherent superposition of mass eigenstates, where the (in)appropriate quantum mechanical treatment affects the neutrino mass values derived from cosmological data \cite{Fuller}.
One commonly notices that the {\em flavor} associated energies are defined through the averaged value from Eq.~(\ref{eq01C}).
We shall demonstrate in the following that the (re)interpretation of the probabilistic concepts for a composite quantum system through the principles of the generalized measurement theory agrees with the assertion that the above definition is inadequate.

\section{Flavor weighted energies}

Supposing that the density matrix of a composite quantum system of two neutrino flavor states is given by
\begin{equation}
\rho\bb{t} \equiv \rho = w_e \, M^{e}_{(t)} + w_{\mu} \, M^{\mu}_{(t)},
\label{eq05}
\end{equation}
with $w_{e} + w_{\mu} = 1$, one easily finds that the re-defined probabilities of measuring the electron and muon flavor eigenstates at time $t$ are given by
\begin{eqnarray}
P^{e}_{(t)}   = \mbox{Tr}\{M^{e}_{(0)}\, \rho\} &=& w_e \, \mbox{Tr}\{M^{e}_{(0)}\,M^{e}_{(t)}\} + w_{\mu} \, \mbox{Tr}\{M^{e}_{(0)}\,M^{\mu}_{(t)}\} = \nonumber\\
                        &=& w_e \, \mathcal{P}_{e\rightarrow e}\bb{t} + w_{\mu} \, \mathcal{P}_{\mu\rightarrow e}\bb{t},\\
P^{\mu}_{(t)} = \mbox{Tr}\{M^{\mu}_{(0)}\, \rho\} &=& w_e \, \mbox{Tr}\{M^{\mu}_{(0)}\,M^{e}_{(t)}\} + w_{\mu} \, \mbox{Tr}\{M^{\mu}_{(0)}\,M^{\mu}_{(t)}\} = \nonumber\\
                        &=& w_e \, \mathcal{P}_{e\rightarrow \mu}\bb{t} + w_{\mu} \, \mathcal{P}_{\mu\rightarrow \mu}\bb{t},
\label{eq06}
\end{eqnarray}
where we have used the results from Eqs.~(\ref{eq02B}-\ref{eq02C}) from which one easily notices that
\begin{eqnarray}
P^{e}_{(t)} +  P^{\mu}_{(t)} &=&
w_{e} (\mathcal{P}_{e\rightarrow e}\bb{t} + \mathcal{P}_{e\rightarrow \mu}\bb{t})
+
w_\mu (\mathcal{P}_{\mu\rightarrow e}\bb{t} + \mathcal{P}_{\mu\rightarrow \mu}\bb{t})
= w_{e} + w_\mu = 1
\label{eq06B}
\end{eqnarray}
and that the properties of a statistical mixture are immediate.
It leads to a reinterpretation of the energy related to each flavor quantum number.

The standard {\em total} averaged energy for a composite quantum system is defined through the density matrix as
\begin{eqnarray}
E_{(t)} = \mbox{Tr}\{H \, \rho\} &=& w_{e} \, \mbox{Tr}\{H \, M^{e}_{(t)}\} + w_{\mu} \, \mbox{Tr}\{H \, M^{\mu}_{(t)}\}\nonumber\\
                            &=&   w_{e} \, E^{e}_{(t)} + w_{\mu} \, E^{\mu}_{(t)},
\label{eq07}
\end{eqnarray}
from which one can notice the explicit dependence on the flavor-{\em averaged} energies, $E^{e,\mu}_{(t)}$, recovered from Eq.~(\ref{eq02}).
In this context $E^{e}_{(t)}$ and $E^{\mu}_{(t)}$ are respectively decoupled from the statistical weights $w_{\mu}$ and $w_{e}$.
It just ratifies our previous arguments that such flavor energies are noway correlated with the flavor probabilities from Eq~(\ref{eq07}), $P^{e}_{(t)}$ and $P^{\mu}_{(t)}$ since both of them depend simultaneously on both statistical weights, $w_{\mu}$ and $w_{e}$.
Thus the arguments that assert the ambiguity and the insufficiency in defining the flavor eigenstate averaged energies through $E^{e,\mu}_{(t)}$ are maintained.

To overcome such incongruities we suggest that some kind of flavor-{\em weighted} energy should be considered in order to
establish a univoque correspondence between flavor eigenstate energies and the statistical definitions of probabilities, $P^{e,\mu}_{(t)}$,
After simple mathematical manipulations involving the definitions from Eq.~(\ref{eq01B}) and the probabilities from Eq.~(\ref{eq06}), one easily finds that
\begin{eqnarray}
M^{\mu}_{(0)} \, \rho\, M^{\mu}_{(0)} &=& (w_e \, \mathcal{P}_{e\rightarrow \mu} \bb{t}+ w_{\mu} \, \mathcal{P}_{\mu\rightarrow \mu}\bb{t}) M^{\mu}_{(0)} = P^{\mu}_{(t)}\, M^{\mu}_{(0)},\nonumber\\
M^{e}_{(0)} \, \rho\, M^{e}_{(0)}     &=& (w_e \, \mathcal{P}_{e\rightarrow e}\bb{t} + w_{\mu} \, \mathcal{P}_{\mu\rightarrow e}\bb{t}) M^{e}_{(0)} = P^{e}_{(t)}\, M^{e}_{(0)},
\label{eq08}
\end{eqnarray}
Observing the cyclic properties of the trace, the flavor-{\em weighted} energies can be defined as
\begin{eqnarray}
\epsilon^{e,\mu}_{(t)} = \mbox{Tr}\{M^{e,\mu}_{(0)}\,H\, M^{e,\mu}_{(0)} \, \rho \} &=& \mbox{Tr}\{H\, M^{e,\mu}_{(0)} \, \rho\, M^{e,\mu}_{(0)}\}
= P^{e,\mu}_{(t)}\mbox{Tr}\,\{M^{e,\mu}_{(0)}\,H\} =
P^{e,\mu}_{(t)}\, E^{e,\mu}_{(0)},
\label{eq09}
\end{eqnarray}
which can be promptly compared with the previous definition through the relation
\begin{equation}
\frac{|\epsilon^{e,\mu}_{(t)} - w_{e,\mu} E^{e,\mu}_{(0)}|}{E^{e,\mu}_{(0)}} = |w_{e} - w_{\mu}| \sin^{\2}\bb{2\theta}
\, \sin^{\2}\left(\frac{\Delta E}{2}\,t\right).
\label{eq10A}
\end{equation}
In allusion to the interference phenomenon in quantum mechanics such a residual term has an interference character since it intrinsically brings simultaneous information of $e$ and $\mu$ flavors.
One should notice that its time-averaged value is not zero, which means that the above analysis lead to different interpretations for the mean values of flavor-{\em averaged} and {\em weighted} energies.

One can also easily identify that the {\em total} averaged energy differs from the sum of flavor-{\em weighted} energies by a {\em residual} energy term given by
\begin{eqnarray}
\xi_{(t)}   &=& E_M - (\epsilon^{e}_{(t)} + \epsilon^{\mu}_{(t)}) \nonumber\\
            &=& \mbox{Tr}\{H\,\rho\} - (\mbox{Tr}\{M^{e}_{(0)}\,H\, M^{e}_{(0)} \, \rho\} + \mbox{Tr}\{M^{\mu}_{(0)}\,H\, M^{\mu}_{(0)} \, \rho\})\nonumber\\
            &=& (\mbox{Tr}\{H\, M^{e}_{(0)} \, \rho\} - \mbox{Tr}\{M^{e}_{(0)}\,H\, M^{e}_{(0)} \, \rho\}) + (\mbox{Tr}\{H\, M^{\mu}_{(0)} \,\rho\} - \mbox{Tr}\{M^{\mu}_{(0)}\,H\, M^{\mu}_{(0)} \, \rho\})\nonumber\\
            &=& \mbox{Tr}\{M^{\mu}_{(0)}\,H\, M^{e}_{(0)} \, \rho\} + \mbox{Tr}\{M^{e}_{(0)}\,H\, M^{\mu}_{(0)} \, \rho\}\nonumber\\
            &=& \mbox{Tr}\{M^{\mu}_{(0)}\,H\, M^{e}_{(0)} \, \rho\} + h.c.
\label{eq10B}
\end{eqnarray}
where we have used the unitarity from $M^{e}_{(0)} + M^{\mu}_{(0)}  = \mathbf{1}$.

To summarize our results up to this point, we introduce the simplificative variables:  $w = w_{e}$ and $\delta w = w_{\mu} - w_{e}$ so that one obtains the simplified expressions,
\begin{eqnarray}
E_M &=& \bar{E} + \frac{\delta w}{2} \Delta E \, \cos(2\theta)\\
\epsilon^{e}_{(t)} + \epsilon^{\mu}_{(t)} &=& \bar{E} + \frac{\delta w}{2} \Delta E \, \cos(2\theta) \left[1 - 2\,\sin^{\2}\left(\frac{\Delta E}{2} \, t \right) \,  \sin^{\2}(2\theta)\right]
\label{eq11}
\end{eqnarray}
for the total energies, from which the flavor-{\em residual} energy results in
\begin{equation}
\xi_{(t)} = \delta w \, \Delta E \, \cos\bb{2\theta} \,  \sin^{\2}\bb{2\theta}
\, \sin^{\2}\left(\frac{\Delta E}{2}t\right),
\label{eq13}
\end{equation}

\begin{figure}
\vspace{-2 cm}
\epsfig{file= 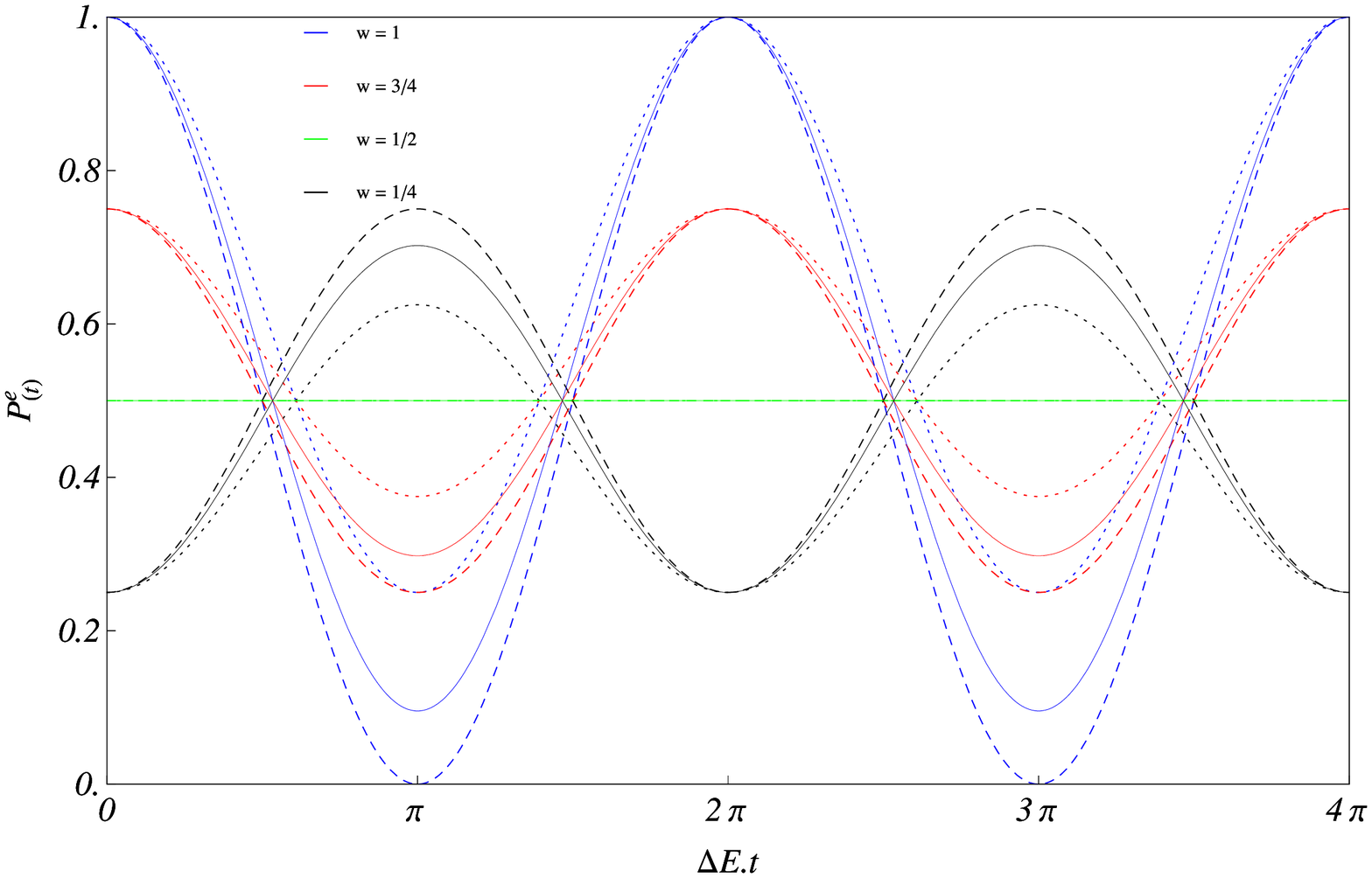, width= 8.5 cm}
\vspace{-0.1cm}
\epsfig{file= 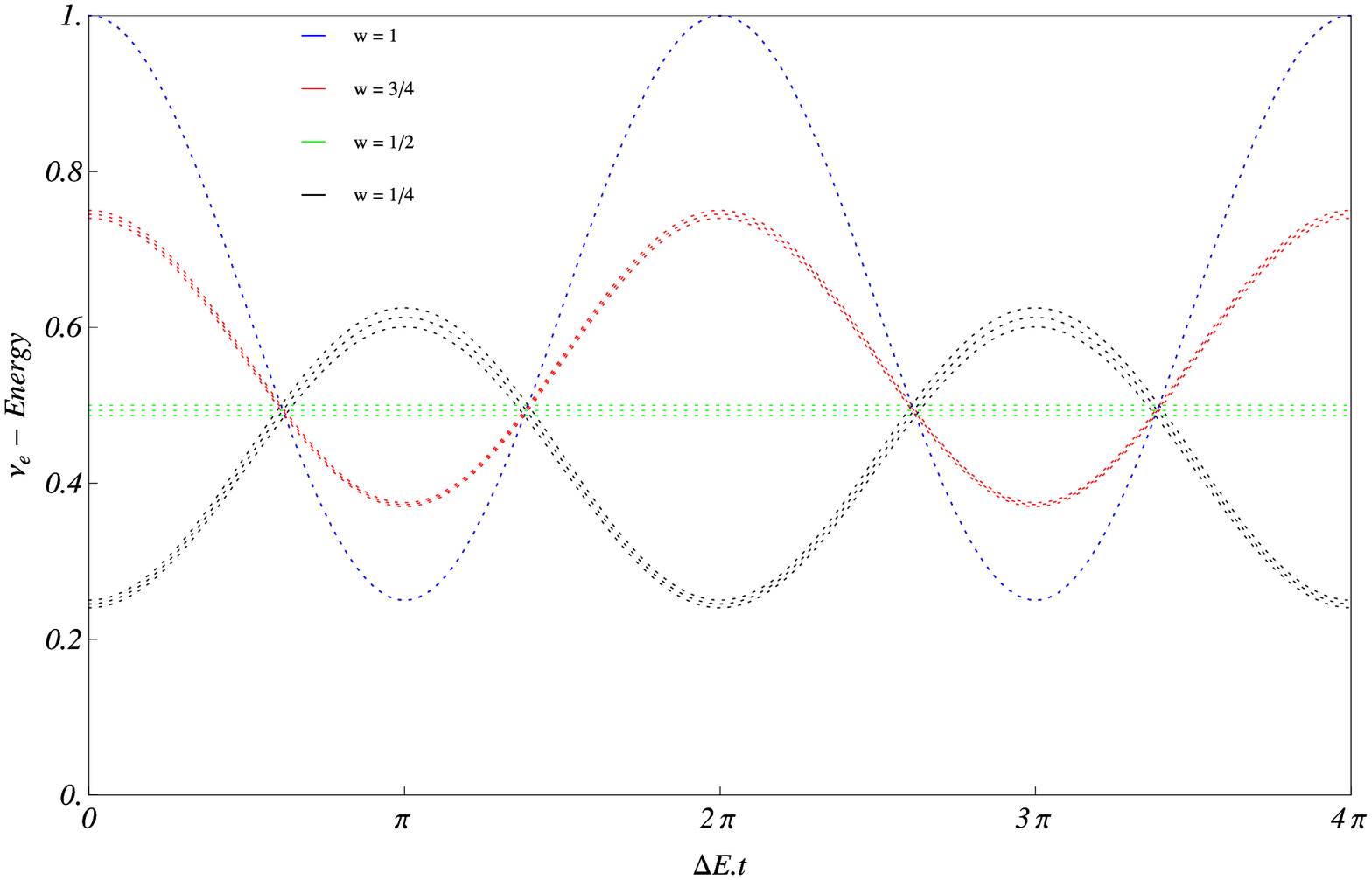, width= 8.5 cm}
\vspace{-0.1cm}
\epsfig{file= 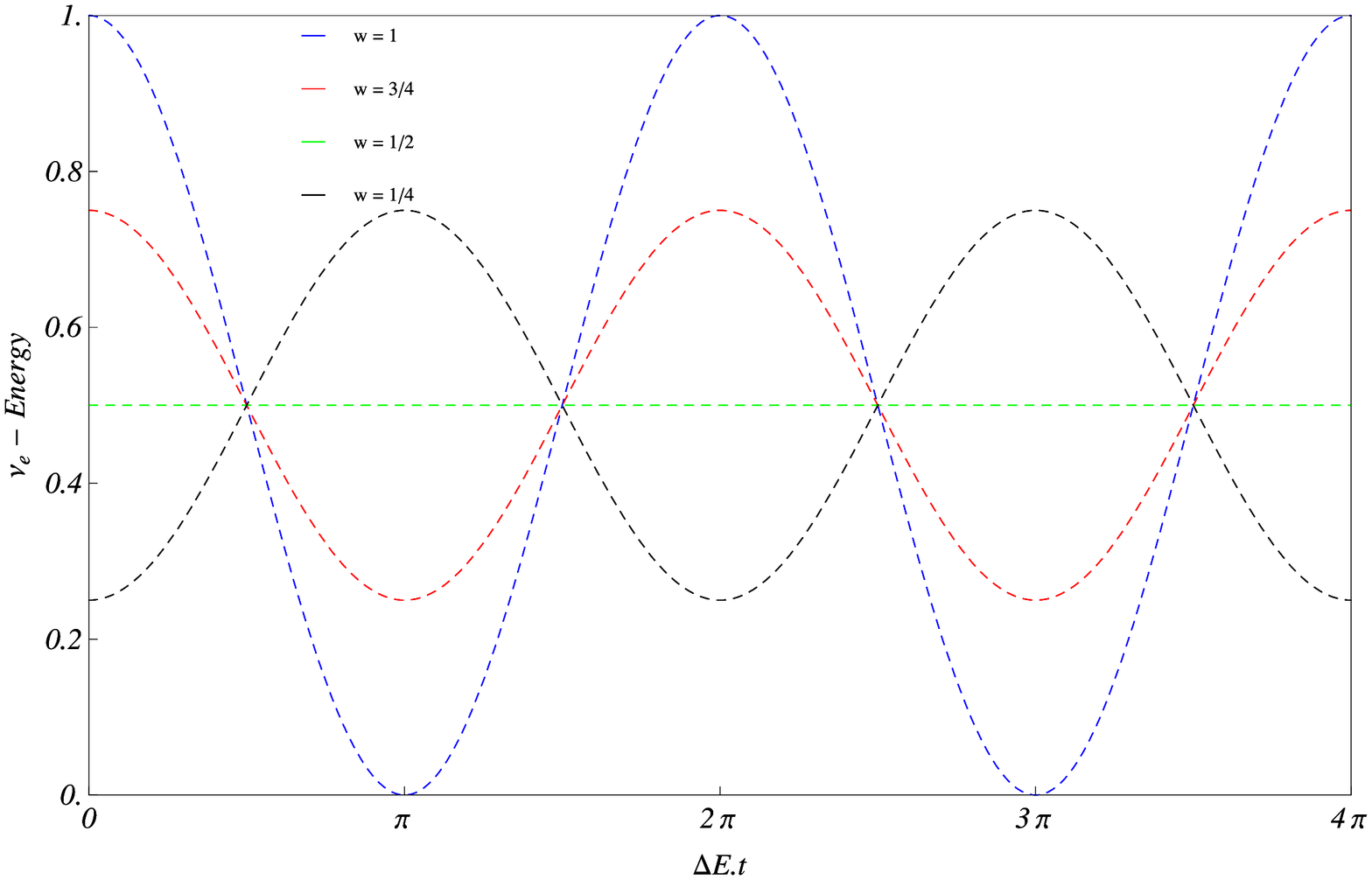, width= 8.5 cm}
\vspace{-0.1cm}
\epsfig{file= 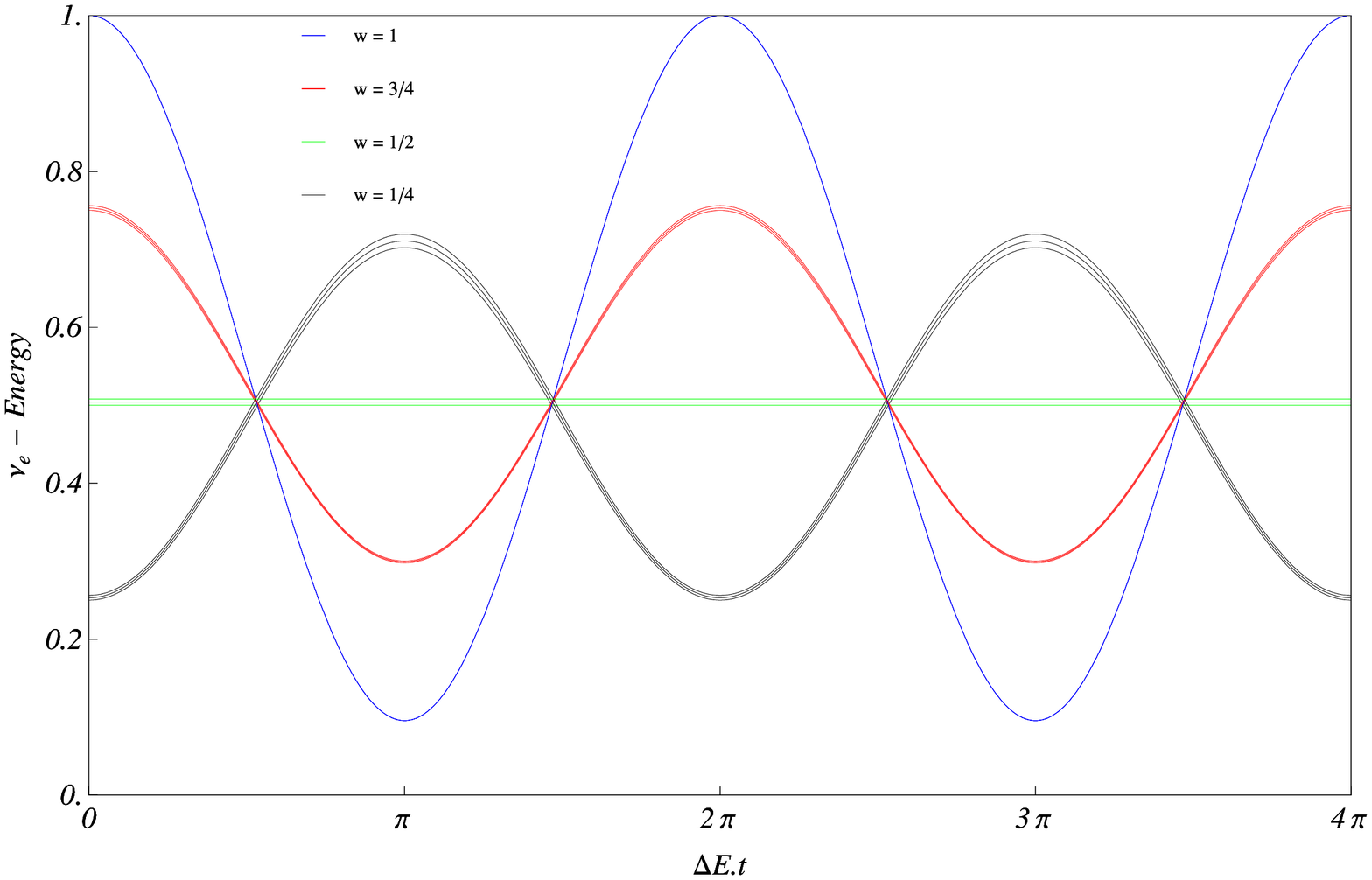, width= 8.5 cm}
\vspace{-0.2cm}
{\small{\caption{\label{an1} Dependence on the oscillating phase $\Delta E\, t$ for the electronic ($e$) flavor oscillating probabilities, $P^{e}_{(t)}$, for several statistical weights ($w = 1,\, 3/4,\, 1/2,$ and $1/4$), in correspondence with the electronic flavor-{\em weighted}  energies, $\epsilon^{e}_{(t)}$ (normalized by the {\em total} averaged energy $E$), for three values of the mixing angle: $\theta = \pi/3$ (dashed lines), $\theta = \pi/4$ (dotted lines),  and $\theta = \pi/5$ (solid lines).
Notice the correspondence between flavor energies and probabilities.
For the maximal statistical mixture ($w = 1/2$) there is no oscillating behavior.
To verify the influence of different relativistic regimes for the propagating mass eigenstates (from NR to UR limits), we have considered $m/p =  10,\, 1,$ and $0.1$ for each plot.
In this case, one can notice some degenerescence effects through three approximated curves by varying $m/p$ from NR to UR limits.}}}
\end{figure}

\begin{figure}
\epsfig{file= 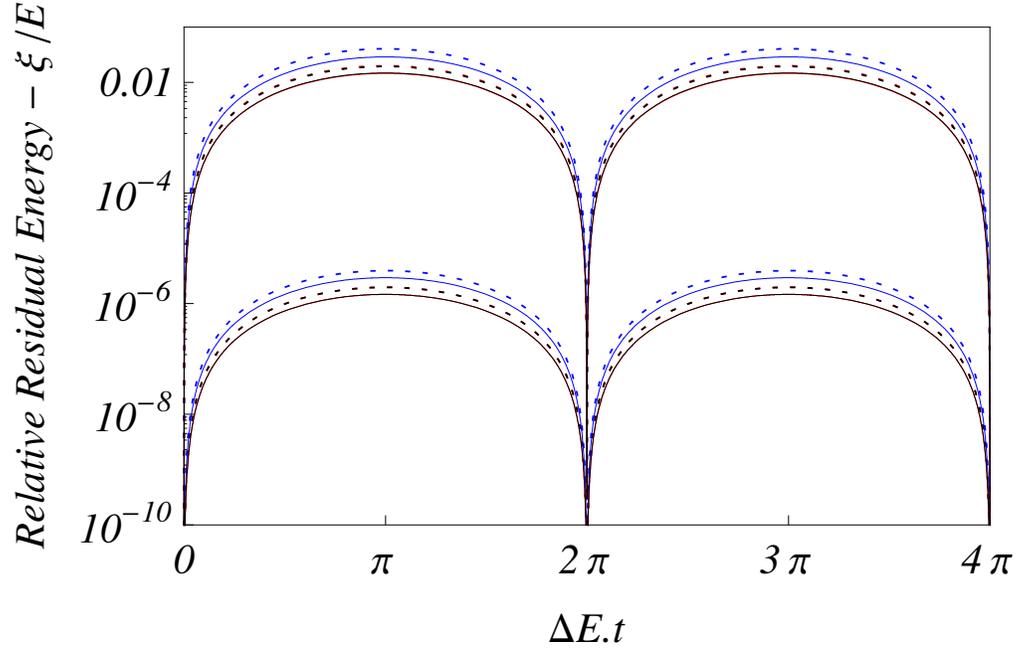, width= 14 cm}
\caption{\label{an2} Relative residual energy $\xi_{(t)} / E$ dependence on the oscillating phase $\Delta E\, t$  ($e$) for several statistical weights ($w = 1,\, 3/4,\, 1/2,$ and $1/4$) in correspondence with the electronic flavor-{\em weighted}  energies, $\epsilon^{e}_{(t)}$ for three values of the mixing angle: $\theta = \pi/3$, $\theta = \pi/4$,  and $\theta = \pi/5$.
We have considered the mixing angles: $\theta = \pi/3$ (dashed lines)  and $\theta = \pi/5$ (solid lines), in correspondence with Fig.~\ref{an1}.
For maximal mixing conditions, i. e.  with $\theta = \pi/4$, the relative residual energy is null.
The same effect is observed for pure states ($w = 1$, or even $w = 0$) and for the maximal statistical mixture ($w = 1/2$).}
\end{figure}

The oscillating probability and the corresponding flavor-{\em weighted} energy for some particular value of the mixing angle, $\theta$, and of the statistical weight, $w$, for different regimes of propagation parameterized by $m/p$, are described in Fig.~\ref{an1}.
For comparative effects, in Fig.~\ref{an2} we discuss the residual energy given by Eq.~(\ref{eq13}).
It is relevant in determining the convergence of our analysis to the standard treatment of quantum mechanics that results in the definition of $E_M$ given by Eq.~(\ref{eq01A}).
The results of Fig.~\ref{an2} ratifies that for pure states ($w = 1$ or $w = 0$) and for the maximal statistical mixture ($\delta w = 0$) the relative residual energy is null.
The same effect is observed when maximal mixing conditions are set by $\theta = \pi/4$.
Depending on the mixing angle, the residual energy $\xi$ can be highly suppressed, as we shall notice in the following section.

Turning back to quantum fundamentals of the above analysis, it is important to emphasize that the definition of flavor-{\em weighted} energies reflects some concepts of the generalized theory of quantum measurements \cite{Breuer,10,20,30}.
There are important variants of quantum measurements schemes that are encountered in practice.
The above results turn out that the generalized measurement theory based on notions of operations and effects is very singular.
The generalized measurement theory leads, in a natural way, to the extended idea of a {\em positive operator-valued measure} which associates with each measurement outcome $\alpha$ a positive operator $M^{\alpha}_{(0)}$.
It may be viewed as an immediate generalization of the von-Neumann-L\"{u}ders projection postulate that introduces the notion of {\em selective} and {\em non-selective} measurements \cite{Breuer}.
At our analysis $\alpha$ corresponds to the quantum numbers related to electronic and muonic flavors, $e$ and $\mu$.

The measurement outcome $\alpha$ represents a classical random number with probability distribution given by Eq.~(\ref{eq06}) where $M^{\alpha}_{(0)}$ is a positive operator called the {\em effect}.
For the case that the measurement is a {\em selective} one, the sub-ensemble of those systems for which the outcome $\alpha$ has been found is to be described by the density matrix
\begin{equation}
\rho_{\alpha} = \left(P^{\alpha}_{(t)}\right)^{-1} \, M^{\alpha}_{(0)}\, \rho \, M^{\alpha}_{(0)}
\label{eq14},
\end{equation}
where $M^{\alpha}_{(0)}\, \rho \, M^{\alpha}_{(0)}$ is called {\em operation}, which maps positive operators into positive operators.
Notice that one consistently has
\begin{equation}
\mbox{Tr}\{\rho_{\alpha}\} = \left(P^{\alpha}_{(t)}\right)^{-1} \,  \mbox{Tr}\{ M^{\alpha}_{(0)}\, \rho \, M^{\alpha}_{(0)}\}
= \left(P^{\alpha}_{(t)}\right)^{-1} \,  \mbox{Tr}\{ M^{\alpha}_{(0)}\, \rho\} = 1.
\label{eq14C}
\end{equation}
For the corresponding {\em non-selective} measurement one has the density matrix
\begin{equation}
\rho^{\prime} = \sum_{\alpha} {P^{\alpha}_{(t)} \rho_{\alpha}},
\label{eq15}
\end{equation}
from which it is also easily verified that $\mbox{Tr}\{\rho^{\prime}\} = 1$.

At our approach, flavor-{\em averaged} energies, $E^{\alpha}_{(0)}$, result from the density matrix for {\em selective} measurements, $\rho_{\alpha}$, from Eq.~(\ref{eq14}), as it can be verified through the relation,
\begin{equation}
E^{\alpha}_{(0)} = \mbox{Tr}\{H \, \rho_{\alpha}\}
\label{eq14B},
\end{equation}
and flavor-{\em weighted} energies, $\epsilon^{\alpha}_{(t)}$, result from the density matrix for {\em non-selective} measurements, $\rho^{\prime}$, from Eq.~(\ref{eq15}), as
\begin{equation}
\sum_{\alpha = e,\mu,\tau}{\epsilon^{\alpha}_{(t)}} = \sum_{\alpha = e,\mu,\tau} {P^{\alpha}_{(t)}\,E^{\alpha}_{(t)}}=  \mbox{Tr}\{H \, \rho^{\prime}\}.
\label{eq15B}
\end{equation}
Each energy component $E^{\alpha}_{(t)}$, with $\alpha = e,\,\mu,\,\tau$, in the above equations are respectively decoupled from its corresponding statistical weight, $w_{\alpha}$ (c. f. Eq.~(\ref{eq14B})).
Therefore, flavor-{\em averaged} energies, $E^{\alpha}_{(t)}$, are noway correlated with the flavor probabilities from Eq~(\ref{eq06}).
The conversion probabilities, $P^{\alpha}_{(t)}$, have multiple dependencies on all statistical weights, $w_{e}$, $w_{\mu}$ and $w_{\tau}$.
The inaccuracy in correlating flavor-{\em averaged} energies, $E^{\alpha}_{(t)}$, with flavor eigenstates is consequently obvious.
Otherwise, the {\em total} averaged energy given by
\begin{eqnarray}
E_{(t)} = \mbox{Tr}\{H \, \rho\} &=& \sum_{\alpha = e,\mu,\tau} w_{\alpha} \, \mbox{Tr}\{H \, M^{\alpha}_{(t)}\} =  \sum_{\alpha = e,\mu,\tau} w_{\alpha} \, E^{\alpha}_{(t)},
\label{eq07BBB}
\end{eqnarray}
seems to be well-defined, in the sense that it is independent of the measurement scheme.
From such a novel interpretation, flavor-{\em weighted} energies are naturally embedded into the quantum measurement scheme and the results can be easily extended to $n$-flavor oscillating quantum systems.

\section{Von-Neumann entropy and quantum measurements}

Now let us report about some basics on quantum statistics and thermodynamics, from which the von-Neumann entropy provides an important entropy functional defined in terms of the density matrix by
\begin{equation}
S\bb{\rho} = - \mbox{Tr}\{\rho\bb{t}\, \ln\bb{\rho\bb{t}}\},
\label{eq26Z}
\end{equation}
where we have set the multiplicative Boltzmann constant, $k_{B}$, equal to unity.
The entropy $S\bb{\rho}$ quantifies the departure of a composite quantum system from a pure state, i. e. it measures the degree of mixture of a state describing a given finite system.
As one can expect, quantum measurements induce modifications on the the von-Neumann entropy of the system.
The entropy change due to a {\em non-selective} measurement scheme described by {\em operations} parameterized by the projection operators $M^{\alpha}_{(0)}$ is given by
\begin{equation}
\Delta S = S\bb{\rho^{\prime}} - S\bb{\rho} \geq 0,
\label{eq26A}
\end{equation}
where
\begin{equation}
S\bb{\rho^{\prime}} =
S\left(\sum_{\alpha} {P^{\alpha}_{(t)} \rho_{\alpha}}\right).
\label{eq26B}
\end{equation}
Since $\Delta S \geq 0$, the {\em non-selective} ideal quantum measurement never decreases the von-Neumann entropy.
An additional property concerns the variation of the entropy involved in the transition from {\em selective} to {\em non-selective} levels of a measurement.
The quantity
\begin{equation}
\delta S =
S\left(\sum_{\alpha} {P^{\alpha}_{(t)} \rho_{\alpha}}\right) - \sum_{\alpha} {P^{\alpha}_{(t)} S\left(\rho_{\alpha}\right)}.
\label{eq26C}
\end{equation}
can be interpreted as a mixing entropy.
It corresponds to the difference between the entropy of a system projected by a {\em non-selective} quantum measurement, $S\left(\sum_{\alpha} {P^{\alpha}_{(t)} \rho_{\alpha}}\right)$, and the average of the entropies of the sub-ensembles, $\rho_{\alpha}$, described by states $M^{\alpha}_{(0)}$.
All the above defined entropies satisfy some set of inequalities \cite{Breuer} which have been extensively used in different forms in the framework of quantum information theory and quantum entanglement.
In particular, one should notice that in case of the previously discussed condition of $\rho_{\alpha} = M^{\alpha}_{(0)}$ for the {\em selective} measurement scheme, with $M^{\alpha}_{(0)}$ denoting the creation of a single-flavor state, the mixing entropy is reduced to
\begin{equation}
\delta S = S\left(\sum_{\alpha} {P^{\alpha}_{(t)} \rho_{\alpha}}\right),
\label{eq26D}
\end{equation}
since $S\left(\rho_{\alpha}\right) = 0$.
It brings up an important meaning to the von-Neumann entropy in the scope of distinguishing measurement procedures.
The {\em non-selective} measurement of flavor-{\em weighted} energies as functionals of the flavor conversion probabilities, modifies the von-Neumann entropy by $\Delta S$ from Eq.~(\ref{eq26A}).
Otherwise, the {\em selective} measurement of flavor-{\em averaged} energies, which are not expressed in terms of flavor conversion probabilities, does not modify the von-Neumann entropy in case of a single-flavor created ensemble.
In case of maximal statistical mixtures, the entropy change due to a {\em non-selective} measurement is null, i. e. $S\bb{\rho} = S\bb{\rho^{\prime}}$.
At the same time, the mixing entropy is maximal and equal to its maximal value, $\delta S = \ln\bb{n}$, in case of a $n$-level system, and the {\em total} averaged energy is reproduced by the generalization of Eq.~(\ref{eq07BBB}) to $n$-flavors.
Thus, in case of a maximal statistical mixing, the {\em non-selective} measurement does not change neither the energy nor the entropy of the system while the {\em selective} measurement changes the entropy.

\begin{figure}
\epsfig{file= 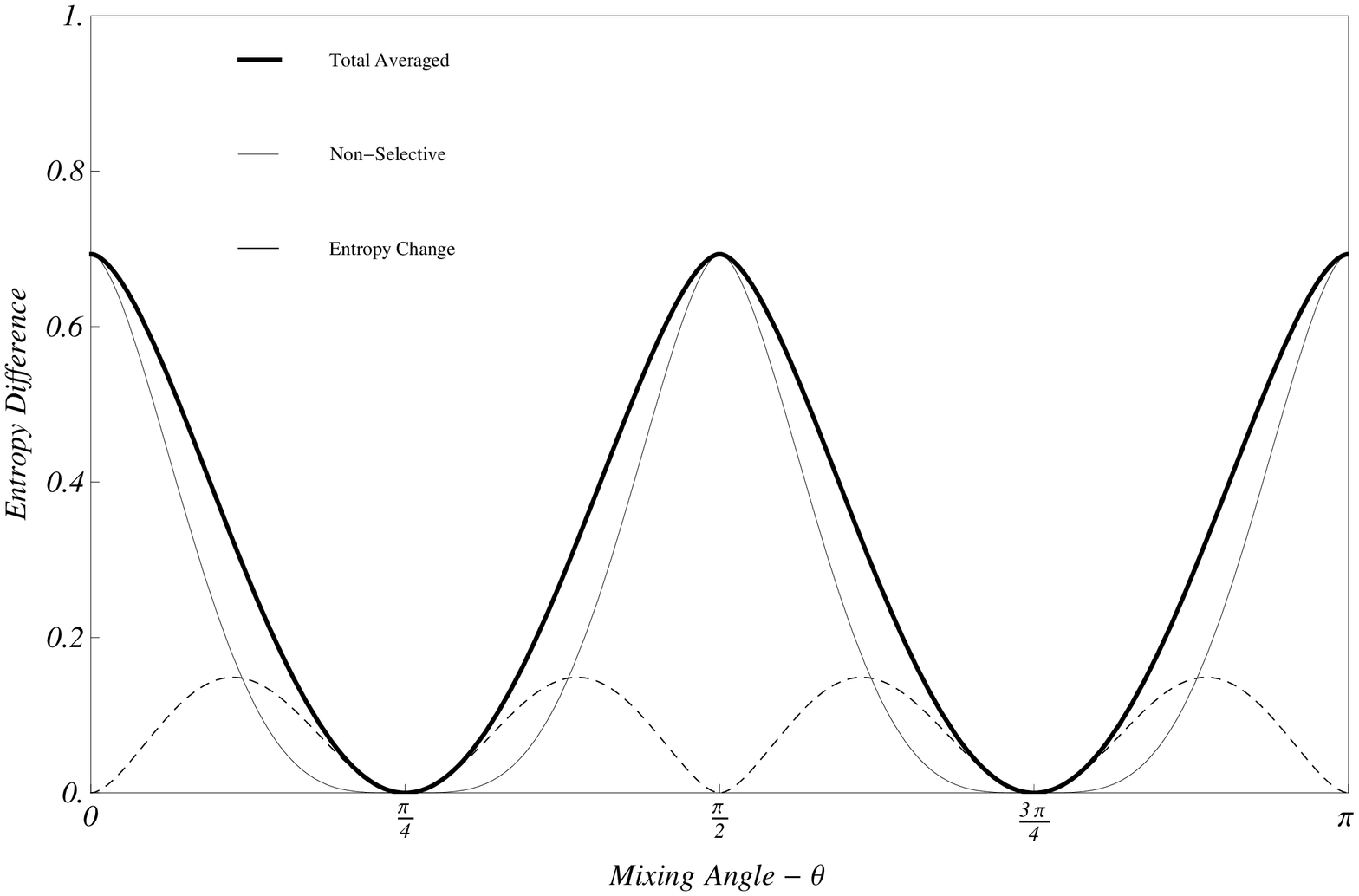, width= 12 cm}
\epsfig{file= 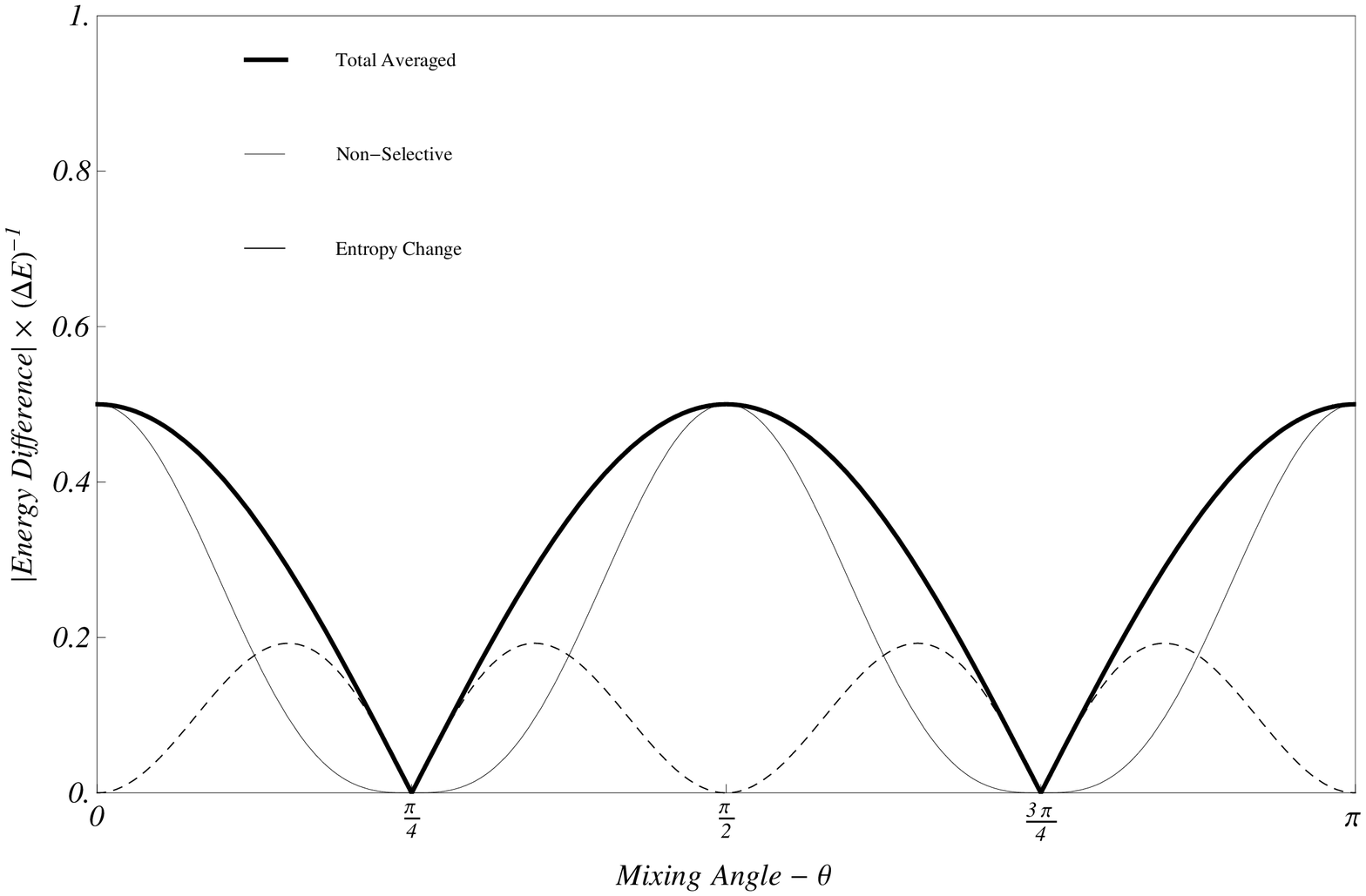, width= 12 cm}
\caption{\label{an4B} Analytical pattern for energy and entropy differences computed through different measurement schemes as function of the mixing angle, $\theta$, for a two-level system.
We have considered the matrix density for a pure state, i. e. when $\delta w = 0$.
The legend is in correspondence with Eq.~(\ref{ABAB}).}
\end{figure}

\begin{figure}
\epsfig{file= 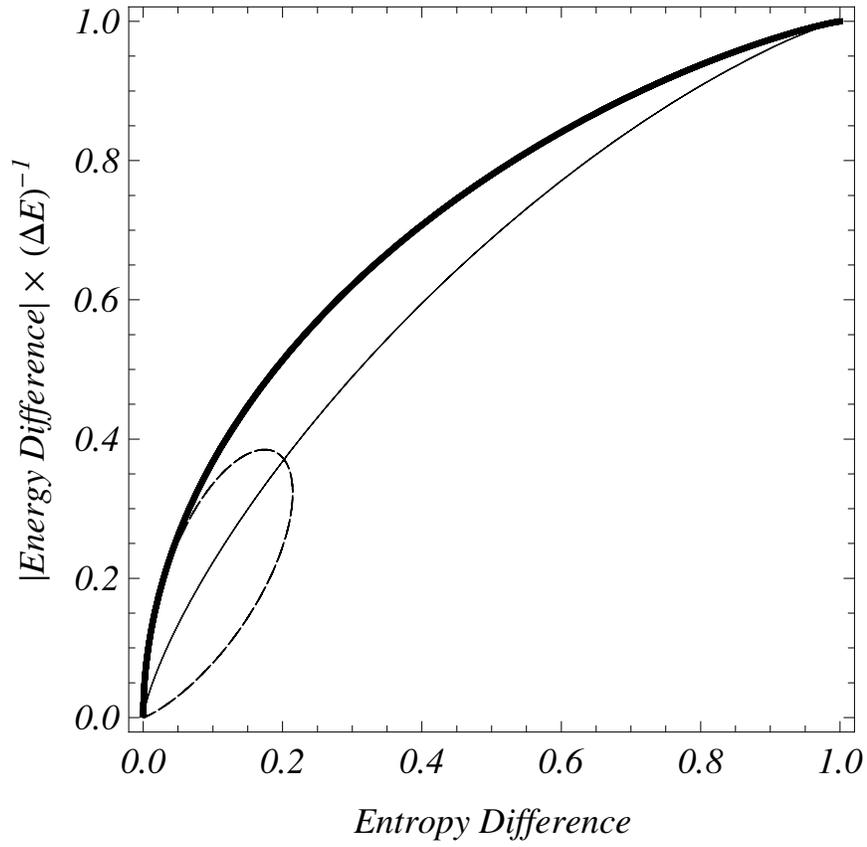, width= 12 cm}
\caption{\label{an5B} Correlation rate between energy and entropy differences (with maximal value normalized to unity).
The curves are in correspondence with the results obtained in Fig.~\ref{an4} when one considers total averaged quantities (thick lines), {\em non-selective} measurements (thin lines) and the corresponding entropy changes (dashed lines).
Is is possible to infer a higher degree of correlation between energy and entropy for the case of {\em non-selective} measurements since it would be maximum for a straight line with angular coefficient equal to unity.}
\end{figure}

By observing the common points between energies and entropies through the above discussed measurement schemes, i. e. by noticing the following correspondence scheme,
\begin{equation}
\begin{array}{llll}
\frac{1}{n}\sum^{n}_{s=1}{E_s} = \bar{E}~~ &\leftrightarrow& ~~ \ln\bb{n} & ~~~~ \mbox{(Maximal Entropy)}\\
\sum_{\alpha} w_{\alpha} E^{\alpha}_{(t)} = E_{(t)} ~~ &\leftrightarrow& ~~ S\bb{\rho} & ~~~~ \mbox{(Total Averaged)}\\
\sum_{\alpha} \epsilon^{\alpha}_{(t)} &\leftrightarrow& ~~ S\bb{\rho^{\prime}} & ~~~~ \mbox{(Non-Selective)}
\end{array}
\nonumber
\end{equation}
it is possible to establish the following correlation between entropy changes and the absolute value of time-averaged energy differences,
\begin{equation}
\begin{array}{llll}
\ln\bb{n} - S\bb{\rho} &\leftrightarrow& ~~ |\langle E_{(t)}\rangle -\bar{E}|& ~~~~ \mbox{(Total Averaged)}\\
\ln\bb{n} - S\bb{\rho^{\prime}} &\leftrightarrow& ~~ | \sum_{\alpha} \langle \epsilon^{\alpha}_{(t)}\rangle -\bar{E}|& ~~~~ \mbox{(Non-Selective)}\\
\Delta S\bb{\rho^{\prime}} &\leftrightarrow& ~~ |\sum_{\alpha} \langle \epsilon^{\alpha}_{(t)}\rangle -\langle E_{(t)}\rangle|& ~~~~ \mbox{(Entropy Change)}\\
\end{array}
\label{ABAB}
\end{equation}
as one can easily depict from Figs.~\ref{an4B} and \ref{an5B} where we have computed time-averaged quantities.
In case of flavor oscillating systems, it corresponds to assuming an intrinsic de-coherence mechanism that suppresses the periodic functions of the density matrix non-diagonal elements.
Such a de-coherence mechanism is equivalent to a delocalization effect that can be achieved by assuming that momentum $(p)$ states weighted by a momentum distribution $(f(p))$ are comprised by an ensemble B in the same way that flavor states are comprised by an ensemble A.
In this case one should have
\begin{equation}
\langle\rho_{AB}\rangle_{t} \equiv Tr_B\{\rho_{AB}\}_{t\rightarrow \infty}
\end{equation}
where $Tr_B\{\}$ denotes an integration over the continuous space of momentum.

Therefore, the entropies computed in terms of $\langle\rho\rangle_{t}$, $S(\langle\rho\rangle_{t})$ can be interpreted as {\em late-time} entropies since one is considering their asymptotic behavior in $t$.
In terms of effective actions, it destroys any coherence behavior of $\rho_{AB}$, which results in a statistical mixing described by $\langle\rho_{AB}\rangle_{t} = \rho_{A}$.

For the two-level system discussed above, from Fig.~\ref{an4B}, one can identify a similar analytical pattern between the energy and entropy time-averaged values in terms of their dependence on the mixing angle, $\theta$.
The Fig.~\ref{an5B} shows a kind of correlation rate between the above quantities when they are normalized by each respective maximum value.
In spite of discussing a two-level system, the qualitative analysis of the results depicted from Fig.~\ref{an5B} allow us to identify, by varying the mixing angle, a higher level of correlation between flavor-{\em weighted} energies and the respective entropies when {\em non-selective} measurements are taken into account.
It simply corresponds to an indication that when one assumes flavor-{\em weighted} energies as the quantifiers for the energy associated to flavor eigenstates, the loss of information due to the measurement procedure is better quantified by the {\em non-selective} related entropy.
To summarize, we shall compute some effective quantities related to flavor-{\em weighted} energies in the following section.

\section{Corrections to the single-particle quantum mechanics of cosmological neutrinos}

The recent issues \cite{Fuller,Kis08,Bel99,Dol02} on quantum mechanics of cosmological neutrinos has focused on finding an appropriate procedure for computing the neutrino mass values derived from cosmological data.
To illustrate an application of our analysis, we reproduce some results from the single-particle quantum mechanics of flavor oscillations reported by Fuller and Kishimoto \cite{Fuller} and we show how the density matrix theory supports such results.

From the Standard Model and cosmological point of views, the main assumption for the cosmological neutrino background is that neutrinos and antineutrinos should be in thermo-chemical equilibrium with the photon- and $e^\pm$-plasma at early times, namely when background temperatures are $T > T_{\rm Dec} \sim 1\,{\rm MeV}$.
For $T\ll T_{\rm Dec}$, the neutrinos and antineutrinos would be completely decoupled, comprising seas of {\em free streaming} particles with energy-momentum and flavor distributions reflecting the equilibrium prior to decoupling, followed by the expansion of the universe.

Assuming that neutrinos are forced by weak interaction-mediated scattering into flavor eigenstates in the pre-decoupling time, the neutrino momentum distribution for each flavor can be approximated by Fermi-Dirac distribution functions, with the number density $\nu_\alpha$'s in a momentum interval $dp$ given by \cite{Fuller}
\begin{equation}
\label{dist}
dn_{\nu_\alpha}= {{1}\over{2\pi^2}}\cdot {{p^2}\over{ e^{E_{\nu_{\alpha}}(a)/ T(a) - \eta_{\nu_\alpha}} +1}}\,dp,
\end{equation}
where we have assumed natural units by setting $\hbar = c = k_B = 1$, and we have reported about the textbook's ratio of chemical potential to temperature for neutrino species $\nu_\alpha$, $\eta_{\nu_\alpha}$.
One can also notice that $T_\nu (a) = T_{\rm Dec} a_{\rm Dec} / a$ is an effective neutrino temperature obtained by assuming that $E_\nu(a) / T_{\nu}(a)$ is a co-moving quantity defined in order to satisfy the Boltzmann equation for equilibrium conditions so that
\begin{equation}
\left.\frac{\partial}{\partial a}\left(\frac{1}{e^{E_\nu(a) / T_{\nu}(a) - \eta_{\nu_\alpha}} + 1}\right)\right|_{T = T_{(equil)}} = 0.
\label{cond}
\end{equation}

In a previous issue \cite{Fuller}, the energy-momentum dispersion relation was introduced in order to give
\begin{equation}
E_\nu (a) \approx T_{\rm Dec} \left( \epsilon^2 + \frac{\sum_i \vert U_{\alpha i} \vert^2 m_i^2}{T_{\rm Dec}^2} \right)^{1/2} .
\end{equation}
that leads to the standard choice for the effective mass of a neutrino in flavor eigenstates $\nu_\alpha$,
\begin{equation}
m^2_{\rm{eff}, \nu_\alpha} = \sum_i \vert U_{\alpha i} \vert^2 m_i^2.
\label{meff}
\end{equation}
It is interpreted as the dynamical mass for ultrarelativistic neutrinos of flavor $\nu_\alpha$.
In particular it has been shown that when the neutrino momentum redshifts forward non-relativistic regimes \cite{ag97}, this effective mass is no longer relevant in characterizing the energy-momentum dispersion relation.
As explicitly pointed out by Fuller and Kishimoto \cite{Fuller}, one can also notice that energy distribution functions for neutrinos in mass eigenstates can be approximated by weighted sums of the flavor eigenstates,
\begin{equation}
d n_{\nu_i} = \sum_\alpha \vert U_{\alpha i} \vert^2 d n_{\nu_\alpha},
\label{eq22AA}
\end{equation}
that has been used to compute the neutrino energy density $\rho_E$ through which one can derive the neutrino mass values after some phenomenology.
The usual textbook \cite{Dodelson} method for computing the energy density of neutrinos in the universe can be rewritten \cite{Fuller} in using the effective mass from Eq.~(\ref{meff}) and the number density distributions of neutrinos in flavor eigenstates from Eq.~(\ref{eq22AA}), so that one has
\begin{equation}
\rho^{(Std)}_E = \sum_\alpha \int ( p^2 + m^2_{\rm{eff}, \nu_\alpha} )^{1/2} d n_{\nu_\alpha} .
\label{naive}
\end{equation}
However, Fuller and Kishimoto \cite{Fuller} assumes that to calculate the energy density of these particles, the mass eigenstate energy could be introduced in order to give
\begin{eqnarray}
\rho^{\nu}_E & = & \sum_i \int ( p^2 + m_i^2 )^{1/2} d n_{\nu_i} \nonumber \\
 & = & \sum_{i, \alpha} \int \vert U_{\alpha i} \vert^2 ( p^2 + m_i^2 )^{1/2} d n_{\nu_\alpha} .
\label{std}
\end{eqnarray}
At equivalently small redshifts corresponding to large scale factors, the neutrino freeze-out regime is intensified and neutrinos become non-relativistic, i. e. the magnitude of the neutrino masses become more relevant than the momentum magnitude at late times.
At early times, the ultra-relativistic regime naturally suppress any eventual divergence from the naive effective mass approach.

Through the above reported single-particle quantum mechanics framework already quantified and discussed by Fuller and Kishimoto \cite{Fuller}, the distribution function used in the Eq.~(\ref{eq22AA}) for mass eigenstates would have the same Fermi-Dirac form only if the degeneracy parameters for the three flavors were all equal.
It obviously follows from the common sense unitarity described by $\sum_\alpha \vert U_{\alpha i} \vert^2 = 1$.
The distribution functions of neutrinos in mass eigenstates would not have a Fermi-Dirac form when the degeneracy parameters were not identical for all three active flavors \cite{Fuller}.

From this point we assume that the energy-momentum dispersion relation related to $E_{\nu}$ has to be defined through the analysis of the previous section, for what the subtleties circumventing the general theory of quantum measurements for composite quantum systems are relevant.
As an additional outstanding result obtained from our analysis, the ambiguities introduced by the confront between the results of Eqs.~(\ref{naive}) and (\ref{std}) will disappear if one considers the flavor-{\em weighted} energy for computing the neutrino energy density $\rho_E$.

Reporting about a phenomenologically consistent analysis that involves three neutrino flavor eigenstates, a generical interpretation can be depicted from our results.
We assume that each of the flavor ensemble is described by a normalized state vector $\nu^{\alpha}$, with $\alpha = e, \, \mu, \, \tau$, in the underlying Hilbert space.
It is then natural to study the statistics of the total ensemble by mixing the flavor ensembles with respective weights $w_{\alpha}$.
The mixing is achieved by taking a large number $N_{\alpha}$ of systems from each flavor ensemble so that $w_{\alpha} = N_{\alpha} / \sum{N_{\alpha}}$.
Thus the maximal statistical mixture with $w_e = w_{\mu} = w_{\tau}$ results from the assumption that $dn_e = dn_{\mu} = dn_{\tau}$.
In this case, the {\em total} averaged energy computed in the previous section would lead to a series of convergent results,
\begin{equation}
E_M = \sum_{\alpha = e,\mu,\tau} w_{\alpha} E^{\alpha}_{(0)} = \sum_{\alpha = e,\mu,\tau} \langle P^{\alpha} \rangle E^{\alpha}_{(0)} = \sum_{\alpha = e,\mu,\tau} \langle \epsilon^{\alpha}\rangle = \frac{1}{3}\sum^{3}_{s=1}{E_s} = \bar{E}.
\label{eq26}
\end{equation}
so that all flavor energy definitions would reproduce exactly the same results for the neutrino energy densities.

Turning back to our simplified two flavor analysis it would be easily verified by setting $\delta w = 0$ in Eqs.~(\ref{eq11}-\ref{eq13}).
Obviously, in the limits of pure states, namely when $w = 0, \, 1$, or even when $\delta w \neq 0$, flavor-{\em weighted} energies lead to different predictions for $\rho_E$.
To clarify this point it is convenient to compute the time-averaged values of the flavor energy definitions that we have introduced.

One should notice that the periodic functions of the oscillating phase, $\sin{\bb{\Delta E\, t}}$ and $\cos{\bb{\Delta E\, t}}$,  for cosmological time scales can be averaged to zero.
Let us then integrate the oscillating phase from the time of decoupling, with scale factor $a = a_{\rm Dec}$, when neutrinos still were ultrarelativistic, up to the present epoch, with $a = 1$, in order to obtain
\begin{equation}
\int \Delta E dt \gtrsim
\frac{\Delta m^{\2}}{2q} \int_{a_{\rm Dec}}^{1} \frac{a}{\dot{a}} da \approx
\frac{\Delta m^{\2}}{2q} \, H_{\0}^{-1} \int_{a_{\rm Dec}}^{1} a^{n} da \approx
\frac{\Delta m^{\2}}{2q} \, H_{\0}^{-1}
\label{eq27}
\end{equation}
where we have considered that $H = H_{\0} \dot{a}/a \sim a^{-n}$, $n > 0$, is the Hubble rate during the remaining time from radiation to matter domination eras, and $q = k_{B} T^{\nu}_{\0}$ is the co-moving momentum.
Maintaining the assumption of natural units and observing that $H_{\0}^{-1} \sim 0.7 \times 10^{33} \,{\rm eV}^{-1}$, $\Delta m^{2} \lesssim 2.4 \times 10^{-3} \,{\rm eV}^{2}$ and $q \sim 0.167 \times 10^{-4}\, {\rm eV}$, one finds a huge oscillation number given by $\Delta E\, \tau \sim 10^{34}$, justifying the time-average ($\langle \rangle_{\rm time}$) procedure.

\begin{figure}
\epsfig{file= 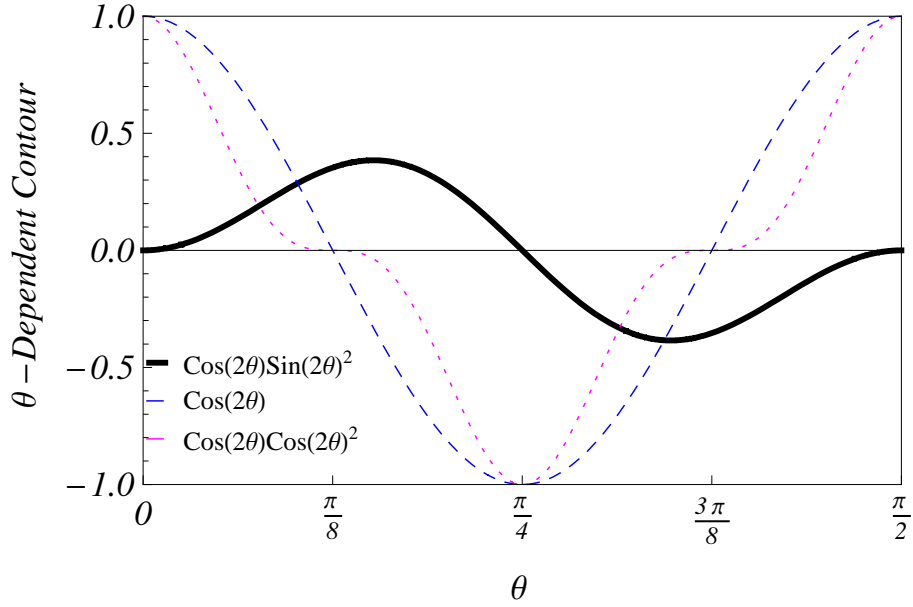, width= 14 cm}
\caption{\label{an3} Contour function for the modifications of the time-averaged energy expressions.
Each curve describes the suppression of the analytical dependence on the statistical weights ($\delta w$) of the {\em total} averaged energy (dashed line), $\langle E_M \rangle$, of the flavor-{\em weighted} energies (solid line), $\langle \epsilon^{e} + \epsilon^{\mu} \rangle$, and of the {\em residual} energy (dotted line), $\langle\xi\rangle$, as functions of the mixing angle.}
\end{figure}

Since $\langle \sin^{\2}\left(\frac{\Delta E}{2} t\right)\rangle_{time} = \frac{1}{2}$, the time-averaged quantities that are relevant to us are
\begin{eqnarray}
\langle E_M \rangle
&=& \bar{E} + \frac{\delta w}{2} \Delta E \, \cos\bb{2\theta},\nonumber\\
\langle \epsilon^{e} + \epsilon^{\mu} \rangle
&=& \bar{E} + \frac{\delta w}{2} \Delta E \, \cos\bb{2\theta} \, \cos^{\2}\bb{2\theta}),\nonumber\\
\langle \xi \rangle
&=& \bar{E} + \frac{\delta w}{2} \Delta E \, \cos\bb{2\theta} \, \sin^{\2}\bb{2\theta}),
\label{eq20}
\end{eqnarray}
which can be conveniently manipulated to obtain
\begin{eqnarray}
(\delta w)^{-1} \left(\langle E_M \rangle - \bar{E}\right)
&=& \frac{\Delta E}{2} \, \cos(2\theta),\nonumber\\
(\delta w)^{-1} \left(\langle \epsilon^{e} + \epsilon^{\mu} \rangle - \bar{E}\right)
&=& \frac{\Delta E}{2}\, \cos(2\theta) \, \cos^{\2}(2\theta),\nonumber\\
(\delta w)^{-1} \langle \xi \rangle
&=& \frac{\Delta E}{2} \, \cos(2\theta) \, \sin^{\2}(2\theta),
\label{eq20C}
\end{eqnarray}
that differ one from each other by the corresponding dependence on the mixing angle, $\theta$, as it is qualitatively illustrated in Fig.~\ref{an3}.
One can easily notice that $\bar{E} = \frac{1}{D}\sum^{D}_{s=1}{E_s}$ is the input into Eq.~(\ref{std}) used to compute the neutrino energy density, $\rho^{\nu}_E$, in case of $D$ mass eigenstates.
Since it is defined in terms of the mass eigenstate eigenvalues for a well-defined Hamiltonian, we consider $\rho^{\nu}_E$ as our reference for the standard quantum mechanical procedure that results in measurable energy densities.
Eq.~(\ref{eq20C}) suggests at least three methods for discussing the fractional difference between energy densities, $\rho_E$.
All of them depend on the difference between the statistical weights, $\delta w$, and on the modulation given by the contour function illustrated in Fig.\ref{an3}.
The common variable among them is the mass-energy difference, $\Delta E$, from which one can identify the following auxiliary variable,
\begin{eqnarray}
\Delta \rho & = & \int \,\Delta E\, d n_{\nu_s},
\label{stdd}
\end{eqnarray}
that with
\begin{eqnarray}
\rho^{\nu}_E & = &\frac{1}{2} \int \,\sum^{2}_{s=1}{E_s} d n_{\nu_s},
\label{stddd}
\end{eqnarray}
allows one to quantify difference among the three predicitions derived from $E_M - \bar{E}$, $(\sum_{\alpha}{E^{\alpha}}) - \bar{E}$ and $(\sum_{\alpha}{E^{\alpha}}) - E_M$,  obtained from {\em total} averaged, $E_M$, flavor-{\em weighted}, $\sum_{\alpha}{E^{\alpha}}$, and the quantum mechanical mass averaged, $\bar{E}$.

\begin{figure}
\epsfig{file= 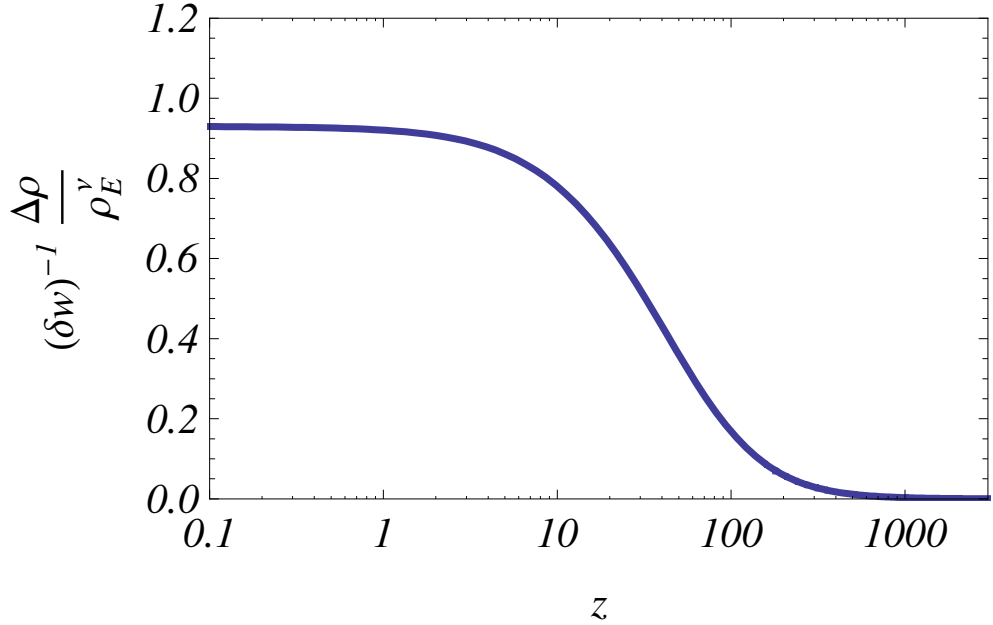, width= 14 cm}
\caption{\label{an4} $\Delta \rho/ \rho^{\nu}_E$ as a function of redshift.
The curve is for $m_{\2} \simeq 10 \, k_{B} T^{\nu}_{\0} = 1.67 ~{\rm meV}$  and  $m_{\1} = \simeq 49 ~{\rm meV}$.}
\end{figure}

\begin{figure}
\epsfig{file= 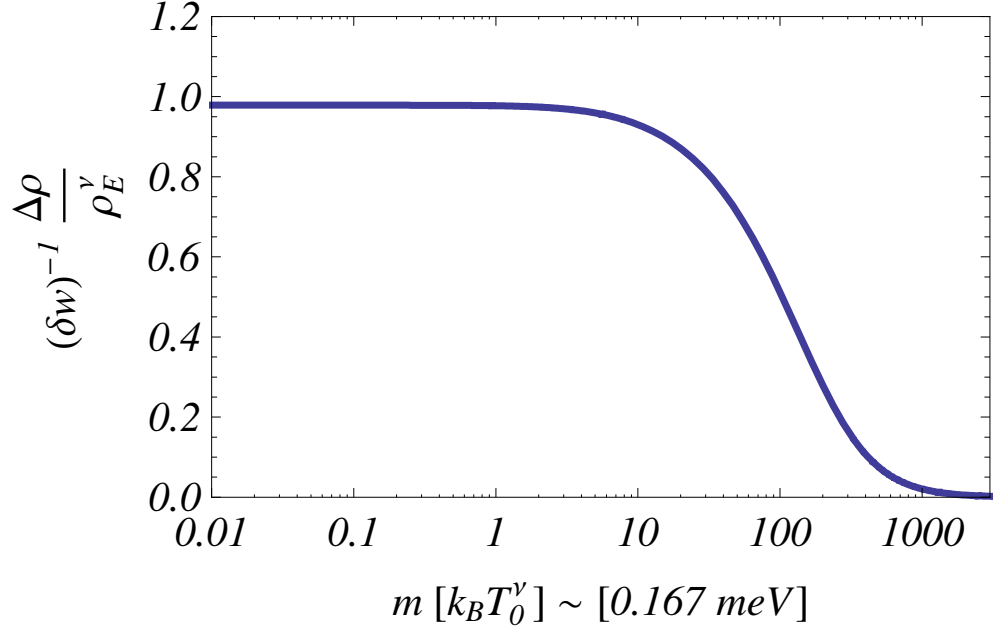, width= 14 cm}
\caption{\label{an5} $\Delta \rho/ \rho^{\nu}_E$ as a function of the smaller mass $m_{\2} = m$  in units of $k_{B} T^{\nu}_{\0} = 0.167 ~{\rm meV}$.
The curve is for $\Delta m^{\2} = m^{\2}_{\1}-m^{\2}_{\2} \simeq 2.4 \times 10^{-3}\, {\rm eV}^{\2}$ at the present epoch ($z = 0$).}
\end{figure}

Figs.\ref{an4} and \ref{an5} show the maximal fractional difference between the energy densities computed through these different techniques.
Besides being suppressed by the eventual null value of $\delta w$, the values depicted from Figs. \ref{an4} and \ref{an5} are modulated by the mixing angle dependent functions of Fig.\ref{an3}.

At early times, neutrino momenta are large enough that all the mass eigenstates are ultrarelativstic and masses have small effects on the total neutrino energy density.
At recent times, the neutrino {\em free streaming} regime is no longer ultrarelativistic and rest mass becomes significant in the energy density composition.
However, we reinforce that all the measurement schemes are equivalent in case of a maximal statistical mixing, i. e. when $\delta w = 0$.
The kinematics of the neutrinos is thus that one reproduced by Eq.(\ref{std}).
Otherwise, if $\delta w \neq 0$, as can be seen from Figs.~\ref{an4} and \ref{an5}, the disparity resulting from different quantum measurement schemes becomes significant.
It has to be taken into account when one quantifies the matter power spectrum, once the dark matter energy density in neutrinos and the character of their kinematics are relevant to the formation of large scale structures.

The results of the above analysis can be immediately extended to a composite system of three flavor eigenstates, i. e. electronic-$e$, muonic-$\mu$, and tauonic-$\tau$ neutrinos, for which the $3 \times 3$ density matrix is given by
\begin{equation}
\rho = w \mathbf{1} - \delta w_{\1} \, M^{\mu}_{(t)} -  \delta w_{\2} \, M^{\tau}_{(t)}
\label{eq23}
\end{equation}
with $w = w_e$, $\delta w_{\1} = w_{e} - w_{\mu}$, $\delta w_{\2} = w_{e} - w_{\tau}$, and $M^{e}_{(t)} + M^{\mu}_{(t)} + M^{\tau}_{(t)} = \mathbf{1}$.
In case of a maximal statistical mixture, the time-averaged flavor probabilities are easily extracted as
\begin{equation}
\langle P^{e,\mu,\tau}\rangle =  \frac{1}{3}
\label{eq24}
\end{equation}
and the corresponding time-averaged flavor-{\em weighted} energies are given by
\begin{eqnarray}
\langle \epsilon^{e} \rangle &=& \frac{\bar{E}}{3} + \frac{1}{36}\left[ 6\, \delta_{\1\2}\, \cos\bb{2\theta_{\1\2}} \, \cos^{\2}\bb{\theta_{\1\3}} - (\delta_{\2\3} - \delta_{\3\1}) (1 - 3 \cos\bb{2\theta_{\1\3}})\right]\nonumber\\
\langle \epsilon^{\mu} \rangle &=& \frac{\bar{E}}{3} + \frac{1}{9}\left[
(\delta_{\3\1} - \delta_{\2\3})
\cos^{\2}\bb{\theta_{\1\3}}\,\sin^{\2}\bb{\theta_{\2\3}}\right.\nonumber\\
&& +
(\delta_{\1\2}-\delta_{\3\1})
\left(\sin\bb{\theta_{\1\2}}\,\cos\bb{\theta_{\2\3}}
+
\cos\bb{\theta_{\1\2}}\,\sin\bb{\theta_{\1\3}}\,\sin\bb{\theta_{\2\3}}\right)^{\2}\nonumber\\
&&\left.+
(\delta_{\2\3} - \delta_{\1\2})
\left(\cos\bb{\theta_{\1\2}}\,\cos\bb{\theta_{\2\3}}
-
\sin\bb{\theta_{\1\2}}\, \sin\bb{\theta_{\1\3}}\,\sin\bb{\theta_{\2\3}}\right)^{\2}
\right]\nonumber\\
\langle \epsilon^{\tau} \rangle &=& \frac{\bar{E}}{3} + \frac{1}{9}\left[
(\delta_{\3\1} - \delta_{\2\3})
\cos^{\2}\bb{\theta_{\1\3}}\,\cos^{\2}\bb{\theta_{\2\3}}\right.\nonumber\\
&& -
(\delta_{\1\2}-\delta_{\3\1})
\left(\cos\bb{\theta_{\1\2}}\, \sin\bb{\theta_{\2\3}} + \sin\bb{\theta_{\1\2}}\,\sin\bb{\theta_{\1\3}}\,\cos\bb{\theta_{\2\3}}\right)^{\2}\nonumber\\
&&\left.-
(\delta_{\2\3} - \delta_{\1\2})
\left(\sin\bb{\theta_{\1\2}}\,\sin\bb{\theta_{\2\3}} - \cos\bb{\theta_{\1\2}}\,\sin\bb{\theta_{\1\3}}\,\cos\bb{\theta_{\2\3}}\right)^{\2}
\right]
\label{eq25}
\end{eqnarray}
where $\delta_{ij} = E_{i} - E_{j}$.
The above obtained results can be summed up in order to verify the Eq.~(\ref{eq26}) and the quantum mechanical definition from Eq.~(\ref{std}), which is equal to the sum of the mass eigenstate energy eigenvalues, $\bar{E}$.
The result can be easily extended to any $D$-dimension composite quantum system when one assumes a maximal statistical mixture.
  
\section{Conclusion}

The generalized framework for constructing the quantum measurement schemes and their relations with flavor associated energies and corresponding von-Neumann entropies related to {\em selective} and {\em non-selective} measurement schemes were investigated in this manuscript.
Our analysis is adequate for describing the approximate methods which measure the spectrum of quantum observables with finite resolution.
That was the case of flavor-{\em weighted} energies that we have defined in the context of composite quantum systems and compared with the well-know {\em averaged} energies.
We have identified such definitions as generalizations of the von-Neumann-L\"{u}ders projection postulate that introduces the notion of {\em selective} and {\em non-selective} measurements \cite{Breuer}.
At our approach, flavor-{\em averaged} energies, $E^{\alpha}$, result from the density matrix for {\em selective} measurements, and flavor-{\em weighted} energies, $\epsilon^{\alpha}$, result from the density matrix for {\em non-selective} measurements, ratifying each respective association with the corresponding measurement scheme \cite{Breuer}.

The consistency of our approach with previous quantum mechanics predictions and its theoretical support provided by the fundamentals of the generalized theory of quantum measurements shows that the correct interpretation of flavor associated energies, and their inherent correlation with von-Neumann entropies, demands for some statistical description of the flavor oscillation problem.

Our analysis was contextualized in the on-going revolutions in observational cosmology where experimental neutrino physics have provided us with fascinating examples of overlaps between distinct frameworks that result in complementary solutions of physical puzzles.
The complete analysis addressing the neutrino mass values through the cosmological phenomenology should involve two additional aspects that were preliminarily introduced here: i) the coherence of flavor eigenstates and ii) the relevance of the von Neumann entropy on the theory of quantum measurements.
We have shown that cosmological neutrinos emerge as a fascinating example where the salient questions concerning the definition of quantum measurements and the role of quantum operations and effects can be addressed and hopefully better understood.

In addition, one could notice that important properties of quantum entropies are used to characterize the information gained in a quantum measurement or even to comprehend the irreversible nature of the quantum dynamics of open quantum systems.
It is particularly relevant when the quantum entropy of the composite system of cosmological neutrinos submitted to a thermal bath is connected with the cosmological entropy $S = (\rho + p)/T$.
Also different forms of quantum entropies have been proven to be equivalent to physical quantifiers for complex systems.
In particular, the single-particle entanglement associated to particle mixing can be expressed in terms of quantum entropies and transition probabilities in flavor oscillations.
Some previous analysis show that the quantum information encoded in the neutrino flavor states can suffer the effects of delocalization so that the single-particle mode entanglement provide one with the quantum information task that works on the operational characterization of systems of elementary particles.
The neutrino flavor from the cosmological background fluid present a natural delocalization due its Fermi-Dirac characterization given by the momentum distribution function.
The relevance of our analysis can be supplied by its subsequent contribution in predicting more accurate values to the cosmological limits for the transfer of the flavor entanglement of the neutrino fluid into that of a single-particle system with decoherent modes.

\begin{acknowledgments}
A. E. B. would like to thank for the financial support from the Brazilian Agencies FAPESP (grant 08/50671-0) and CNPq (grant 300233/2010-8).
\end{acknowledgments}

\renewcommand{\baselinestretch}{1}

\pagebreak
\newpage

\pagebreak
\newpage

\pagebreak
\newpage

\pagebreak
\newpage

\pagebreak
\newpage

\pagebreak
\newpage

\pagebreak
\newpage
\end{document}